\documentclass[aps,paper]{revtex4}%
\usepackage{amsfonts}
\usepackage{amsmath}
\usepackage{amssymb}
\usepackage{graphicx}%
\begin{document}
\title[$T$-matrix for the short-range perturbation]{Quasirelativistic electronic $T$-matrix for the short-range perturbation}
\author{Sergey A. Ktitorov}
\affiliation{Ioffe Physico-Technical Institute, the Russian Academy of Sciences, St.
Petersburg, 194021, Russia }
\author{Vicco Tamaris}
\affiliation{Williams College, Williamstown, Massachusetts, USA}

\begin{abstract}
Exact $T$-matrix for the delta-function short-range perturbation of the
(3+1)-Dirac equation has been derived. Separability of the potential in the
angular momentum representation is used. A characteristic equation for the
$T$-matric poles determining the bound and resonance states has been obtained.
The calculated $T$-matrix determines transport properties of the narrow-gap
and zero-gap semiconductors.

\end{abstract}
\maketitle









\section{Introduction}

Transport theory and theory of the electronic spectrum need in
calculation of the $T$-matrix including its "nonphysical" part
outside the mass shell. This makes it necessary to consider
exactly solvable models of perturbation. One of them is the delta
function potential. In the case of the two-band nonrelativistic
problem described by the Dirac equation the problem of bound and
resonance states was considered with a use of such potential in
the paper \ref{tamarshortrange}. The short radius potential
$\delta\left( r-r_{0}\right)  $ was considered. This potential has
no a singularity at $r=0$ and is separable in the angular momentum
representation. Here were consider the problem of the bound and
resonance states with slightly different approach and calculate an
exact $T$-matrix. We take into account possible difference of the
perturbation matrix elements calculated on wave functions of the
upper and lower bands that is equivalent to consideration of both
potential and mass perturbations.

\section{Characteristic equation}

As it is well known, the electronic states near the band edge are determined
by the quasirelativistic Dirac equation \cite{tamarshortrange}. It can be
written in the momentum representation in the form:%

\begin{align}
s\mathbf{\sigma p}\chi\left(  \mathbf{p}\right)  +\left(  ms^{2}-E\right)
\varphi\left(  \mathbf{p}\right)  +\int d^{3}p^{\prime}V_{1}\left(
\mathbf{p}-\mathbf{p}^{\prime}\right)  \varphi\left(  \mathbf{p}^{\prime
}\right)   &  =0,\label{diracup}\\
s\mathbf{\sigma p}\varphi\left(  \mathbf{p}\right)  -\left(  ms^{2}+E\right)
\chi\left(  \mathbf{p}\right)  +\int d^{3}p^{\prime}V_{2}\left(
\mathbf{p}-\mathbf{p}^{\prime}\right)  \chi\left(  \mathbf{p}^{\prime}\right)
&  =0, \label{diracdown}%
\end{align}
where $\mathbf{\sigma}$ is the Pauli matrix, $s$ is the quasirelativistic
limiting velocity of band electrons, $2ms^{2}=E_{g}.$ The potential Fourier
transform $V_{i}\left(  \mathbf{p}\right)  $ can be expanded into a series of
the Legendre polynomials $P_{l}\left(  \cos\theta\right)  $ in the case of the
spherical symmetry \cite{Baz}:%
\begin{align}
V_{i}\left(  \left\vert \mathbf{p}-\mathbf{p}^{\prime}\right\vert \right)   &
=\sum_{l}V_{i}^{l}\left(  p,p^{\prime}\right)  P_{l}\left(  \cos\theta\right)
\left(  2l+1\right)  /4\pi,\label{legendre1}\\
V_{i}^{l}\left(  p,p^{\prime}\right)   &  =2\int_{0}^{\infty}drrV_{i}\left(
r\right)  \frac{J_{l+1/2}\left(  pr\right)  J_{l+1/2}\left(  p^{\prime
}r\right)  }{\sqrt{pp^{\prime}}}, \label{legendrebessel}%
\end{align}
where $J_{l+1/2}\left(  pr\right)  $ is the Bessel function. The wave
functions can be expanded into a series of spherical spinors $\Omega
_{jlm}\left(  \mathbf{n}\right)  $ \cite{BLP}:%
\begin{equation}
\left(
\begin{array}
[c]{c}%
\varphi_{E}^{\varkappa m}\left(  \mathbf{p}\right) \\
\chi_{E}^{\varkappa m}\left(  \mathbf{p}\right)
\end{array}
\right)  =\left(
\begin{array}
[c]{c}%
f_{E}\left(  p\right)  i^{-l}\Omega_{jlm}\left(  \mathbf{n}\right) \\
-g_{E}\left(  p\right)  i^{-l^{\prime}}\Omega_{jl^{\prime}m}\left(
\mathbf{n}\right)
\end{array}
\right)  , \label{spinorexpan}%
\end{equation}
where $l=j\mp1/2,$ $l^{\prime}=j\pm1/2,$ $j$ is the total angular momentum
quantum number, $\mathbf{n}=\mathbf{p/}p,$ $\theta$ is the angle between the
vectors $\mathbf{p}$\ and $\mathbf{p}^{\prime},$ while%
\begin{equation}
\mathbf{\sigma n}\Omega_{jlm}\left(  \mathbf{n}\right)  =i^{l^{\prime}%
-l}\Omega_{jl^{\prime}m}\left(  \mathbf{n}\right)  . \label{spiral}%
\end{equation}
Signs $\pm$ correspond to $\varkappa=\pm\left(  j+1/2\right)  .$ Thus, eqs.
(\ref{diracup}), (\ref{diracdown}) take the form%

\begin{equation}
\left(  ms^{2}-E\right)  f_{E}\left(  p\right)  \Omega_{jlm}\left(
\mathbf{n}\right)  -spg_{E}\left(  p\right)  \left(  -1\right)  ^{l-l^{\prime
}}\Omega_{jlm}\left(  \mathbf{n}\right)  +\int d^{3}p^{\prime}\sum
_{l_{1},m_{1}}V_{1}^{l_{1}}\left(  p,p^{\prime}\right)  P_{l_{1}}\left(
\cos\theta\right)  f_{E}\left(  p^{\prime}\right)  \Omega_{jl_{1}m_{1}}\left(
\mathbf{n}^{\prime}\right)  =0, \label{diracupspher}%
\end{equation}

\begin{equation}
\left(  ms^{2}+E\right)  g_{E}\left(  p\right)  \Omega_{jl^{\prime}m}\left(
\mathbf{n}\right)  +spf_{E}\left(  p\right)  \left(  -1\right)  ^{l^{\prime
}-l}\Omega_{jl^{\prime}m}\left(  \mathbf{n}\right)  -\int d^{3}p^{\prime}%
\sum_{l_{1},m_{1}}V_{2}^{l_{1}}\left(  p,p^{\prime}\right)  P_{l_{1}}\left(
\cos\theta\right)  g_{E}\left(  p^{\prime}\right)  \Omega_{jl_{1}^{\prime
}m_{1}}\left(  \mathbf{n}^{\prime}\right)  =0. \label{diracdownspher}%
\end{equation}
Making use of the addition theorem for the spherical functions%
\begin{equation}
\sum_{m}Y_{lm}^{\ast}\left(  \mathbf{n}\right)  Y_{lm}\left(  \mathbf{n}%
^{\prime}\right)  =\frac{2l+1}{4\pi}P_{l}\left(  \cos\theta\right)  ,
\label{addition}%
\end{equation}
of the spherical spinors orthogonality \cite{berest}%
\begin{equation}
\int do\Omega_{jlm}^{\ast}\left(  \mathbf{n}\right)  \Omega_{j_{1}l_{1}m_{1}%
}\left(  \mathbf{n}\right)  =\delta_{j,j_{1}}\delta_{l,l_{1}}\delta_{m,m_{1}}
\label{sphericalortho}%
\end{equation}%
\begin{equation}
\Omega_{j=l\pm1/2,l,m}\left(  \mathbf{n}\right)  =\left(
\begin{array}
[c]{c}%
\pm\sqrt{\frac{j+1/2\pm\left(  m-1/2\right)  }{2\left(  j+1/2\mp1/2\right)  }%
}Y_{lm-1/2}\left(  \mathbf{n}\right) \\
\sqrt{\frac{j+1/2\mp\left(  m+1/2\right)  }{2\left(  j+1/2\mp1/2\right)  }%
}Y_{l^{\prime}m-1/2}\left(  \mathbf{n}\right)
\end{array}
\right)  \label{sphera}%
\end{equation}
we obtain from (\ref{diracupcylinder}), (\ref{diracdowncylynder})%
\begin{align}
\left(  ms^{2}-E\right)  f_{E}\left(  p\right)  +spg_{E}\left(  p\right)
+\int_{0}^{\infty}dp^{\prime}\left(  p^{\prime}\right)  ^{2}f_{E}\left(
p^{\prime}\right)  V_{1}^{l}\left(  p,p^{\prime}\right)   &  =0,\nonumber\\
\left(  ms^{2}+E\right)  f_{E}\left(  p\right)  +spg_{E}\left(  p\right)
-\int_{0}^{\infty}dp^{\prime}\left(  p^{\prime}\right)  ^{2}g_{E}\left(
p^{\prime}\right)  V_{2}^{l^{\prime}}\left(  p,p^{\prime}\right)   &  =0,
\label{lpeqs}%
\end{align}
where $l-l^{\prime}=\pm1.$

$\ $Zero-radius \cite{ostrov} and separable potentials \cite{newton} are
popular in the nonrelativistic scattering theory.\ However, the Dirac equation
is extremely sensitive to a singularity of the potential \cite{zeld}. The
singularity of the delta function potential can be regulated putting the delta
function to the spherical shell \cite{tamarshortrange}:%
\begin{equation}
V_{i}\left(  r\right)  =\frac{V_{i}^{0}}{4\pi r_{0}^{2}}\delta(r-r_{0}).
\label{delta}%
\end{equation}

Substituting (\ref{drlta}) into (\ref{legendrebessel}) we obtain the separable
in the angular momentum-momentum modulus representation potential:%
\begin{equation}
V_{i}^{l^{\prime}}\left(  p,p^{\prime}\right)  =v_{i}^{l}\left(  p\right)
v_{i}^{l}\left(  p^{\prime}\right)  , \label{separable}%
\end{equation}
where%

\begin{equation}
v_{i}^{l}\left(  p\right)  =\sqrt{\frac{V_{i}^{0}}{2\pi pr_{0}}}%
J_{l+1/2}(pr_{0}). \label{factor}%
\end{equation}

Equations (\ref{lpeqs}) become degenerate and can be written as follows:%
\begin{align}
\left(  ms^{2}-E\right)  f_{E}\left(  p\right)  +spg_{E}\left(  p\right)
+v_{1}^{l}\left(  p\right)  \int_{0}^{\infty}dp^{\prime}\left(  p^{\prime
}\right)  ^{2}f_{E}\left(  p^{\prime}\right)  v_{1}^{l}\left(  p^{\prime
}\right)   &  =0,\nonumber\\
\left(  ms^{2}+E\right)  f_{E}\left(  p\right)  +spg_{E}\left(  p\right)
-v_{2}^{l}\left(  p\right)  \int_{0}^{\infty}dp^{\prime}\left(  p^{\prime
}\right)  ^{2}g_{E}\left(  p^{\prime}\right)  v_{2}^{l}\left(  p^{\prime
}\right)   &  =0. \label{separableeqs}%
\end{align}

Introducing the functions%
\[
F\left(  E\right)  =\int_{0}^{\infty}dpp^{2}f_{E}\left(  p\right)  v_{1}%
^{l}\left(  p\right)  ,\text{ \ \ }G\left(  E\right)  =\int_{0}^{\infty
}dpp^{2}g_{E}\left(  p\right)  v_{2}^{l}\left(  p\right)  ,\text{
\ \ }R\left(  p\right)  =\left(  s^{2}p^{2}+m^{2}s^{4}-E^{2}\right)  ^{-1},
\]

we obtain the algebraic equation set%
\begin{align}
F  &  =G\int_{0}^{\infty}dpp^{2}sR\left(  p\right)  v_{1}^{l}\left(  p\right)
v_{2}^{l^{\prime}}\left(  p\right)  -\left(  E+ms^{2}\right)  F\int
_{0}^{\infty}dpp^{2}R\left(  p\right)  \left(  v_{1}^{l}\left(  p\right)
\right)  ^{2},\nonumber\\
G  &  =F\int_{0}^{\infty}dpp^{2}sR\left(  p\right)  v_{1}^{l}\left(  p\right)
v_{2}^{l^{\prime}}\left(  p\right)  +\left(  E-ms^{2}\right)  G\int
_{0}^{\infty}dpp^{2}R\left(  p\right)  \left(  v_{2}^{l^{\prime}}\left(
p\right)  \right)  ^{2}. \label{eqset}%
\end{align}

The solvability condition for this equation set gives the characteristic
equation:%
\begin{align}
\left[  1+\frac{E+ms^{2}}{\hbar^{2}s^{2}}\frac{V_{1}^{0}}{4\pi r_{0}}%
I_{l+1/2}\left(  \kappa r_{0}\right)  K_{l+1/2}\left(  \kappa r_{0}\right)
\right]  \left[  1+\frac{E-ms^{2}}{\hbar^{2}s^{2}}\frac{V_{2}^{0}}{4\pi r_{0}%
}I_{l^{\prime}+1/2}\left(  \kappa r_{0}\right)  K_{l^{\prime}+1/2}\left(
\kappa r_{0}\right)  \right]   &  =\nonumber\\
&  \frac{V_{1}^{0}V_{2}^{0}}{16\pi^{2}r_{0}^{4}\hbar^{2}s^{2}},
\label{character}%
\end{align}

where $\kappa^{2}=\left(  m^{2}s^{4}-E^{2}\right)  \hbar^{-2}s^{-2},$
$I_{n}\left(  x\right)  ,$ $K_{n}\left(  x\right)  $ are the modified Bessel functions.

\section{Calculation of the $T$-matrix}

Let us determine the standard representation bispinor basis as follows%
\begin{equation}
\left(  \mathbf{r,}g\left\vert p,\chi,+\right.  \right)  =\frac{i^{l}}%
{\sqrt{2E_{p}}}\left(
\begin{array}
[c]{c}%
\sqrt{E_{p}+m}R_{pl}\left(  r\right)  \Omega_{\varkappa}\left(  \mathbf{r}%
/r\right) \\
-\sqrt{E_{p}-m}R_{pl^{\prime}}\left(  r\right)  \Omega_{\varkappa^{\prime}%
}\left(  \mathbf{r}/r\right)
\end{array}
\right)  , \label{basis+}%
\end{equation}

\begin{equation}
\left(  \mathbf{r,}g\left\vert p,\chi,-\right.  \right)  =\frac{i^{l^{\prime}%
}}{\sqrt{2E_{p}}}\left(
\begin{array}
[c]{c}%
\sqrt{E_{p}-m}R_{pl}\left(  r\right)  \Omega_{\varkappa}\left(  \mathbf{r}%
/r\right) \\
\sqrt{E_{p}+m}R_{pl^{\prime}}\left(  r\right)  \Omega_{\varkappa^{\prime}%
}\left(  \mathbf{r}/r\right)
\end{array}
\right)  , \label{basis-}%
\end{equation}

where $\lambda=\pm$, $\varkappa=\left(  j,l,m\right)  $; $E_{p}=\sqrt
{m^{2}s^{4}+p^{2}s^{2}}$; $l-l^{\prime}=sgn$ $\varkappa;$ $\varkappa=\pm1,$
$\pm2,$ $\ldots;$ $R_{pl}\left(  r\right)  =\sqrt{p/r}H_{l+1/2}^{\left(
1\right)  }\left(  pr\right)  ;$ $g$ is the bispinor index$.$ We work in units
with $\hbar=s=1.$ The exact wave function of the out-basis satisfies the
Lippmann-Schwinger equation:%
\begin{equation}
\left\langle \mathbf{r,}g\left\vert p,\chi,\lambda\right.  \right\rangle
=\left(  \mathbf{r,}g\left\vert p,\chi,\lambda\right.  \right)  -\sum
_{g^{\prime},g^{\prime\prime}}\int d^{3}rG_{g,g^{\prime}}^{r}\left(
\mathbf{r-r}^{\prime}\right)  V_{g^{\prime},g^{\prime\prime}}\left(
r^{\prime}\right)  \left\langle \mathbf{r}^{\prime}\mathbf{,}g^{\prime\prime
}\left\vert p,\chi,\lambda\right.  \right\rangle \label{lippmann}%
\end{equation}

The perturbation matrix can be written in the form%
\begin{equation}
V_{gg^{\prime}}\left(  r\right)  =V_{g}\delta_{gg^{\prime}}. \label{perturb}%
\end{equation}

Distinct values for the upper and lower bands stem from different symmetries
of the Kohn-Luttenger basic functions. We define $V_{1}=V_{2}=V_{+};$
$V_{3}=V_{4}=V_{-}.$ The $T$- matrix can be defined os follows:%
\begin{equation}
\left\langle p^{\prime},\chi^{\prime},\lambda^{\prime}\right\vert T\left\vert
p,\chi,\lambda\right\rangle \equiv\sum_{g}\int d^{3}r\left(  p^{\prime
}\mathbf{,}\chi^{\prime},\lambda g\left\vert \mathbf{r},g\right.  \right)
V_{g}\left(  \mathbf{r}\right)  \left\langle \mathbf{r,}g\left\vert
p,\chi,\lambda\right.  \right\rangle . \label{Tdef}%
\end{equation}

Using (\ref{lippmann}) and expanding the Green function into a series of the
Dirac Hamiltonian eigenfunctions we obtain the equation for the $T$- matrix:%
\begin{align}
\left\langle p^{\prime},\chi^{\prime},\lambda^{\prime}\right\vert T\left(
E\right)  \left\vert p,\chi,\lambda\right\rangle  &  =\left(  p^{\prime}%
,\chi^{\prime},\lambda^{\prime}\right\vert V\left(  r\right)  \left\vert
p,\chi,\lambda\right)  -\nonumber\\
&  -\int d^{3}p_{1}\sum_{\chi_{1},\lambda_{1}}\frac{\left(  p^{\prime}%
,\chi^{\prime},\lambda^{\prime}\right\vert V\left(  r\right)  \left\vert
p_{1},\chi_{1},\lambda_{1}\right)  }{E_{p_{1}}sgn\lambda_{1}-E-i0}\left\langle
p_{1},\chi_{1},\lambda_{1}\right\vert T\left(  E\right)  \left\vert
p,\chi,\lambda\right\rangle . \label{Teq}%
\end{align}

The perturbation matrix elements are determined as follows%
\begin{equation}
\left(  p^{\prime},\chi^{\prime},\lambda^{\prime}\right\vert V\left(
r\right)  \left\vert p,\chi,\lambda\right)  \equiv\sum_{g}\int d^{3}r\left(
p^{\prime}\mathbf{,}\chi^{\prime},\lambda g\left\vert \mathbf{r},g\right.
\right)  V_{g}\left(  \mathbf{r,}g\left\vert p,\chi,\lambda\right.  \right)
\label{matrixelements}%
\end{equation}

This matrix takes a form of a sum of the two factorized expressions in our
case:%
\begin{align}
\left(  p^{\prime},\chi^{\prime},\lambda^{\prime}\right\vert V\left(
r\right)  \left\vert p,\chi,\lambda\right)   &  =\int do\left[  \left(
p^{\prime},\chi^{\prime},\lambda^{\prime}\right\vert \left.  r_{0}%
,\mathbf{n},+\right)  \left(  r_{0},\mathbf{n},+\right\vert \left.
p,\chi,\lambda\right)  \right]  V_{+}+\nonumber\\
&  \left(  p^{\prime},\chi^{\prime},\lambda^{\prime}\right\vert \left.
r_{0},\mathbf{n},-\right)  \left(  r_{0},\mathbf{n},-\right\vert \left.
p,\chi,\lambda\right)  V_{-}. \label{matrixspecial}%
\end{align}

We seek a solution of the Lippmann-Schwinger equation in the form:%
\begin{equation}
\left\langle \mathbf{r,}g\left\vert p,\chi,+\right.  \right\rangle
=\frac{i^{l}}{\sqrt{2E_{p}}}\left(
\begin{array}
[c]{c}%
\sqrt{E_{p}+m}R_{pl}\left(  r\right)  F_{pl}^{\left(  +\right)  }\left(
r\right)  \Omega_{\varkappa}\left(  \mathbf{r}/r\right) \\
-\sqrt{E_{p}-m}R_{pl^{\prime}}\left(  r\right)  G_{pl^{\prime}}^{\left(
+\right)  }\left(  r\right)  \Omega_{\varkappa^{\prime}}\left(  \mathbf{r}%
/r\right)
\end{array}
\right)  , \label{solutionlippmann+}%
\end{equation}

\begin{equation}
\left\langle \mathbf{r,}g\left\vert p,\chi,-\right.  \right\rangle
=\frac{i^{l^{\prime}}}{\sqrt{2E_{p}}}\left(
\begin{array}
[c]{c}%
\sqrt{E_{p}-m}R_{pl}\left(  r\right)  F_{pl}^{\left(  -\right)  }\left(
r\right)  \Omega_{\varkappa}\left(  \mathbf{r}/r\right) \\
\sqrt{E_{p}+m}R_{pl^{\prime}}\left(  r\right)  G_{pl^{\prime}}^{\left(
-\right)  }\left(  r\right)  \Omega_{\varkappa^{\prime}}\left(  \mathbf{r}%
/r\right)
\end{array}
\right)  . \label{solutionlippmann-}%
\end{equation}

The $T$-matrix can be expressed in terms of the $F_{_{pl}}^{\pm}$ and
$G_{_{pl}}^{^{\pm}}$ functions as follows:%

\begin{align}
\left\langle p_{1},\chi_{1},+\right\vert T\left(  E\right)  \left\vert
p,\chi,+\right\rangle  &  =\frac{\delta_{\chi\chi^{\prime}}}{2\sqrt
{E_{p}E_{p^{\prime}}}}\left[  \sqrt{\left(  E_{p_{1}}+m\right)  \left(
E_{p}+m\right)  }V_{+}^{0}F_{pl_{1}}^{\left(  +\right)  }R_{p_{1}l_{1}}^{\ast
}\left(  r\right)  R_{pl_{1}^{\prime}}\left(  r\right)  +\right. \nonumber\\
&  \left.  \sqrt{\left(  E_{p_{1}}-m\right)  \left(  E_{p}-m\right)  }\right]
V_{-}^{0}G_{pl_{1}^{\prime}}^{\left(  +\right)  }R_{p_{1}l_{1}}^{\ast}\left(
r\right)  R_{pl_{1}^{\prime}}\left(  r\right)  , \label{T++}%
\end{align}

\begin{align}
\left\langle p_{1},\chi_{1},-\right\vert T\left(  E\right)  \left\vert
p,\chi,-\right\rangle  &  =\frac{\delta_{\chi\chi^{\prime}}}{2\sqrt
{E_{p}E_{p^{\prime}}}}\left[  \sqrt{\left(  E_{p_{1}}-m\right)  \left(
E_{p}-m\right)  }V_{+}^{0}F_{pl_{1}}^{\left(  -\right)  }R_{p_{1}l_{1}}^{\ast
}\left(  r\right)  R_{pl_{1}^{{}}}\left(  r\right)  +\right. \nonumber\\
&  \left.  \sqrt{\left(  E_{p_{1}}+m\right)  \left(  E_{p}+m\right)  }%
V_{-}^{0}G_{pl_{1}^{\prime}}^{\left(  -\right)  }R_{p_{1}l_{1}^{\prime}}%
^{\ast}\left(  r\right)  R_{pl_{1}^{\prime}}\left(  r\right)  \right]  ,
\label{T--}%
\end{align}

\begin{align}
\left\langle p_{1},\chi_{1},-\right\vert T\left(  E\right)  \left\vert
p,\chi,+\right\rangle  &  =\frac{\delta_{\chi\chi^{\prime}}\left(  -1\right)
^{j+l^{\prime}}}{2\sqrt{E_{p}E_{p^{\prime}}}}\left[  \sqrt{\left(  E_{p_{1}%
}-m\right)  \left(  E_{p}+m\right)  }V_{+}^{0}F_{pl_{1}}^{\left(  +\right)
}R_{p_{1}l_{1}}^{\ast}\left(  r\right)  R_{pl_{1}}\left(  r\right)  -\right.
\nonumber\\
&  \left.  \sqrt{\left(  E_{p_{1}}+m\right)  \left(  E_{p}-m\right)  }%
V_{-}^{0}G_{pl_{1}^{\prime}}^{\left(  +\right)  }R_{p_{1}l_{1}^{\prime}}%
^{\ast}\left(  r\right)  R_{pl_{1}}\left(  r\right)  \right]  , \label{T-+}%
\end{align}

\begin{align}
\left\langle p_{1},\chi_{1},+\right\vert T\left(  E\right)  \left\vert
p,\chi,-\right\rangle  &  =\frac{\delta_{\chi\chi^{\prime}}\left(  -1\right)
^{j+l}}{2\sqrt{E_{p}E_{p^{\prime}}}}\left[  \sqrt{\left(  E_{p_{1}}+m\right)
\left(  E_{p}-m\right)  }V_{+}^{0}F_{pl_{1}}^{\left(  -\right)  }R_{p_{1}%
l_{1}}^{\ast}\left(  r\right)  R_{pl_{1}}\left(  r\right)  -\right.
\nonumber\\
&  \left.  \sqrt{\left(  E_{p_{1}}-m\right)  \left(  E_{p}+m\right)  }%
V_{-}^{0}G_{pl_{1}^{\prime}}^{\left(  -\right)  }R_{p_{1}l_{1}^{\prime}}%
^{\ast}\left(  r\right)  R_{pl_{1}}\left(  r\right)  \right]  , \label{T+-}%
\end{align}

Substituting (\ref{T++}), (\ref{T--}), (\ref{T-+}) and (\ref{T+-})
into (\ref{Teq}) we obtain two independent sets of equations for
the pairs of
functions $F^{(+)},$ $G^{(+)}$ and $F^{(-)},$ $G^{(-)}:$%
\begin{equation}
\widehat{A}^{\left(  \pm\right)  }\left(
\begin{array}
[c]{c}%
F_{pl_{1}}^{\left(  \pm\right)  }\\
G_{pl_{1}^{\prime}}^{\left(  \pm\right)  }%
\end{array}
\right)  =\left(
\begin{array}
[c]{c}%
c_{p_{1}p}^{\pm\pm}V_{+}^{0}b_{p_{1}p}^{\left(  ++\right)  }+c_{p_{1}p}%
^{\mp\mp}V_{-}^{0}b_{p_{1}p}^{\left(  --\right)  }\\
c_{p_{1}p}^{\mp\pm}V_{+}^{0}b_{p_{1}p}^{\left(  ++\right)  }-c_{p_{1}p}%
^{\pm\mp}V_{-}^{0}b_{p_{1}p}^{\left(  --\right)  }%
\end{array}
\right)  , \label{eqsetFG}%
\end{equation}

where
\begin{equation}
c_{p_{1}p}^{\pm\pm}=\sqrt{\left(  E_{p_{1}}\pm m\right)  \left(  E_{p}\pm
m\right)  },\text{ }c_{p_{1}p}^{\pm\mp}=\sqrt{\left(  E_{p_{1}}\pm m\right)
\left(  E_{p}\mp m\right)  }, \label{c}%
\end{equation}

\begin{equation}
b_{p_{1}p}^{\left(  ++\right)  }=R_{p_{1}l_{1}}^{\ast}R_{pl_{1}},\text{
}b_{p_{1}p}^{\left(  --\right)  }=R_{p_{1}l_{1}^{\prime}}^{\ast}%
R_{pl_{1}^{\prime}},\text{ }b_{p_{1}p}^{\left(  +-\right)  }=pR_{p_{1}l_{1}%
}^{\ast}R_{pl_{1}^{\prime}},\text{ }b_{p_{1}p}^{\left(  -+\right)  }%
=pR_{p_{1}l_{1}^{\prime}}^{\ast}R_{pl_{1}}. \label{b}%
\end{equation}

Matrix elements of $\widehat{A}$ read%
\begin{align}
A_{11}^{\pm}  &  =V_{+}^{0}c_{pp_{1}}^{\pm\pm}b_{p_{1}p}^{\left(  ++\right)
}\left[  1+V_{+}^{0}\left(  E+m\right)  B^{++}\left(  E\right)  \right]
\pm\frac{V_{+}^{0}V_{-}^{0}}{p}c_{pp_{1}}^{\pm\mp}b_{p_{1}p}^{\left(
-+\right)  }B^{\pm\mp}\left(  E\right)  ,\nonumber\\
A_{22}^{\pm}  &  =-V_{-}^{0}c_{pp_{1}}^{\mp\pm}b_{p_{1}p}^{\left(  --\right)
}\left[  1+V_{-}^{0}\left(  E-m\right)  B^{--}\left(  E\right)  \right]
\pm\frac{V_{+}^{0}V_{-}^{0}}{p}c_{pp_{1}}^{\mp\mp}b_{p_{1}p}^{\left(
+-\right)  }B^{\pm\mp}\left(  E\right)  ,\nonumber\\
A_{12}^{\pm}  &  =V_{-}^{0}c_{pp_{1}}^{\mp\mp}b_{p_{1}p}^{\left(  --\right)
}\left[  1+V_{-}^{0}\left(  E-m\right)  B^{--}\left(  E\right)  \right]
\pm\frac{V_{+}^{0}V_{-}^{0}}{p}c_{pp_{1}}^{\mp\pm}b_{p_{1}p}^{\left(
+-\right)  }B^{\pm\mp}\left(  E\right)  ,\nonumber\\
A_{21}^{\pm}  &  =V_{+}^{0}c_{pp_{1}}^{\pm\mp}b_{p_{1}p}^{\left(  ++\right)
}\left[  1+V_{+}^{0}\left(  E+m\right)  B^{++}\left(  E\right)  \right]
\mp\frac{V_{+}^{0}V_{-}^{0}}{p}c_{pp_{1}}^{\pm\pm}b_{p_{1}p}^{\left(
-+\right)  }B^{\mp\pm}\left(  E\right)  , \label{A}%
\end{align}

where the matrix $\widehat{B}\left(  E\right)  $ is determined as follows%
\begin{equation}
\widehat{B}\left(  E\right)  =\int_{0}^{\infty}dp\frac{\widehat{b_{pp}}}%
{E_{p}^{2}-E^{2}-i0}. \label{B}%
\end{equation}

Equating the matrix $\widehat{A}$ determinant $d$ to zero, we obtain the
characteristic equation, which was derived in \cite{we} and in the Section II
using a different approach:%
\begin{align}
d\left(  E\right)   &  \equiv\det\widehat{A}\equiv\left[  1+V_{+}^{0}\left(
E+m\right)  \int_{0}^{\infty}dp\frac{\left\vert R_{pl_{1}}\right\vert ^{2}%
}{E_{p}^{2}-E^{2}-i0}\right]  \cdot\nonumber\\
&  \left[  1+V_{-}^{0}\left(  E-m\right)  \int_{0}^{\infty}dp\frac{\left\vert
R_{pl_{1}^{\prime}}\right\vert ^{2}}{E_{p}^{2}-E^{2}-i0}\right]  -\\
-V_{+}^{0}V_{-}^{0}\int_{0}^{\infty}dp\frac{pR_{pl_{1}^{\prime}}^{\ast
}R_{pl_{1}}}{E_{p}^{2}-E^{2}-i0}\int_{0}^{\infty}dp^{\prime}\frac{p^{\prime
}R_{p^{\prime}l_{1}}^{\ast}R_{p^{\prime}l_{1}^{\prime}}}{E_{p}^{2}-E^{2}-i0}
&  =0. \label{det}%
\end{align}

Solving the inhomogeneous equation set (\ref{eqsetFG}) we obtain an expression
for the $T$-matrix:%
\begin{equation}
\left\langle p_{1},\chi_{1},\lambda_{1}\right\vert T\left(  E\right)
\left\vert p,\chi,\lambda\right\rangle =\frac{\delta_{\chi\chi_{1}}}%
{2d\sqrt{E_{p}E_{p_{1}}}}t_{\lambda_{1}\lambda}\left(  E\right)  ,
\label{Tsolution}%
\end{equation}

where the matrix elements of $\widehat{t}$ are determined as follows%
\begin{align}
\left\langle \pm\right\vert t\left\vert \pm\right\rangle  &  =c^{\pm\pm}%
V_{+}^{0}b_{p_{1}p}^{\left(  ++\right)  }\left[  1+V_{-}^{0}\left(
E-m\right)  B^{--}\right]  \mp c^{\mp\pm}V_{+}^{0}V_{-}^{0}\frac{b_{p_{1}%
p}^{\left(  +-\right)  }}{p}B^{\pm\mp}+\label{tmatrix1}\\
&  V_{-}^{0}c^{\mp\pm}\left[  1+V_{+}^{0}\left(  E+m\right)  B^{++}\right]
\mp c^{\pm\mp}V_{+}^{0}V_{-}^{0}\frac{b_{p_{1}p}^{\left(  -+\right)  }}%
{p}B^{\mp\pm},
\end{align}

\begin{equation}
\left\langle \pm\right\vert t\left\vert \mp\right\rangle =\left(  -1\right)
^{l\left(  l^{\prime}\right)  }i^{l+l^{\prime}}\left\{
\begin{array}
[c]{c}%
c^{\mp\pm}V_{+}^{0}b_{p_{1}p}^{\left(  ++\right)  }\left[  1+V_{-}^{0}\left(
E-m\right)  B^{--}\right]  \pm c^{++}V_{+}^{0}V_{-}^{0}\frac{b_{p_{1}%
p}^{\left(  +-\right)  }}{p}B^{\mp\pm}-\\
V_{-}^{0}c^{\pm\mp}\left[  1+V_{+}^{0}\left(  E+m\right)  B^{++}\right]  \mp
c^{\mp\mp}V_{+}^{0}V_{-}^{0}\frac{b_{p_{1}p}^{\left(  -+\right)  }}{p}%
B^{\pm\mp}.
\end{array}
\right\}  \label{tmatrix2}%
\end{equation}

The formulae (\ref{Tsolution}), (\ref{tmatrix1}), and (\ref{tmatrix2}) give an
exact solution for the $T$-matrix for arbitrary $p$, $p_{1}$, and $E$. Taking
these values on the mass shell $p=p_{1},$ $E=\sqrt{m^{2}+p^{2}},$ we obtain
the scattering matrix.

\section*{Conclusion}

Non-relativistic problem of the electronic spectrum and scattering
described by the Dirac equation is considered in the case of the
two-component short-range perturbation. An exact $T$-matrix both
on- and off-shell has been calculated.


\begin{thebibliography}{9}                                                                                                %


\bibitem {tamarshortrange}V.I. Tamarchenko, S.A. Ktitorov, Soviet Physics --
Solid State, \textbf{19}, 2970 (1977).

\bibitem {Baz}A.I. Baz, Ya.B. Zeldovich and A.M.Perelomov,\textit{Scattering,
reactions and decays in nonrelativistic quantum mechanicsQuantum
Electrodynamics}, Nauka, Moscow, 1971.

\bibitem {BLP}V.B. Berestetskii, E.M. Lifshitz, L.P. Pitaevskii, \textit{Relativistic
quantum theory}, Nauka, Moscow, 1968.

\bibitem {berest}A.I. Akhiezer, V.B. Berestetskii, \textit{Quantum
Electrodynamics}, Nauka, Moscow, 1969.

\bibitem {ostrov}Yu.N. Demkov, V.N. Ostrovskii, \textit{Zero-radius Potential
in Atomic Physics}, Leningradskii University Publishing House, Leningrad, 1975.

\bibitem {newton}Newton
\end{thebibliography}
\end{document}